\documentclass{epl}
\usepackage{graphicx}
\usepackage{dcolumn}
\usepackage{bm}
\usepackage{amssymb}
\usepackage{amsmath}
\newcommand{\degree}{\char'27\kern-.3em\hbox{C}}

\title{Diffusion-induced spontaneous pattern formation on gelation surfaces}

\author{Hiroaki Katsuragi}
\institute{Department of Applied Science for Electronics and Materials, Interdisciplinary Graduate School of Engineering Sciences, Kyushu University, 6-1 Kasugakoen, Kasuga, Fukuoka 816-8580, Japan
}


\begin{document}

\maketitle

\begin{abstract}
Although the pattern formation on polymer gels has been considered as a result of the mechanical instability due to the volume phase transition, we found a macroscopic surface pattern formation not caused by the mechanical instability. It develops on gelation surfaces, and we consider the reaction-diffusion dynamics mainly induces a surface instability during polymerization. Random and straight stripe patterns were observed, depending on gelation conditions. We found the scaling relation between the characteristic wavelength and the gelation time. This scaling is consistent with the reaction-diffusion dynamics and would be a first step to reveal the gelation pattern formation dynamics. 
\end{abstract}

The macroscopic dissipative pattern formation has attracted considerable interest in several recent decades \cite{Cross1,Walgraef1}. Many kinds of soft matters have shown fascinating patterns. In particular, polymer gels create various surface patterns, which are typical non-equilibrium phenomena, under the volume phase transition \cite{Tanaka1}. Because the coupling of phase transition and pattern formation is interesting, the dynamics of this pattern formation has been extensively studied both in terms of experiments and theories \cite{Tokita1,Onuki0}. The mechanical instability due to swelling or shrinking of polymer gels plays a crucial role in such pattern formations. Obviously, this phenomenon does not result from diffusion dynamics. On the other hand, reaction-diffusion systems can also create dissipative patterns called Turing patterns \cite{Turing1}. For example, Quyang and Swinney presented some Turing patterns using the CIMA (Chlorite, Iodide, and Malonic Acid) reaction \cite{Quyang1}. They used gels as a reaction base. It should be noticed that gels did not deform in their experiment. As mentioned above, gels and reaction-diffusion systems can create dissipative patterns independently. To our best knowledge, however, there is no report on the macroscopic pattern formation concerning both of them simultaneously, {\it i.e.}, the pattern formation of gels themselves by the reaction-diffusion dynamics has not been discovered. In this paper we show such a pattern formation that we found on gelation surfaces. 

The behaviors of the pattern wavelength and the diffusion length are discussed through this investigation. It will be an evidence for the reaction-diffusion dynamics. We used the scaling analysis methods to reveal the characteristics of the pattern formation. In general, the scaling analysis methods are very useful to understand the polymer dynamics \cite{deGennes1}. Moreover, they have been applied to many other physical phenomena \cite{Meakin1}. We report the unified scaling relation of gelation surface patterns in this paper. We guess the scaling relation would become helpful to clarify the dynamics of general gelation pattern formation since the scaling relation usually has a certain universality.

\begin{figure}
\begin{center}
\scalebox{0.6}[0.6]{\includegraphics{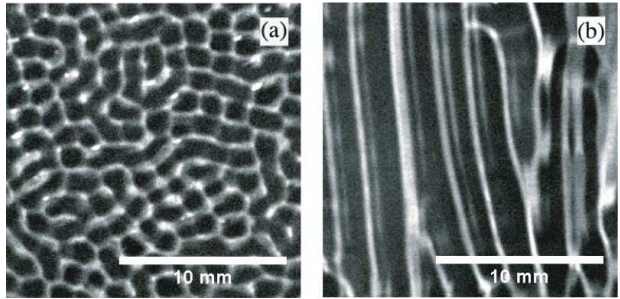}}
\caption{Typical surface patterns of gels: (a)random pattern ($T=36$ {\degree} and $[APS]=0.67$ $g/l$), and (b)straight stripe pattern ($T=50$ {\degree} and $[APS]=0.42$ $g/l$). Bright places correspond to valleys of surface rough patterns. }
\label{fig:pat}
\end{center}
\end{figure}

Poly-acrylamide gels are used in this research. Acrylamide monomer (AA, $M_w = 71.08$) constitutes sub-chains, and methylenbisacrylamide (BIS, $M_w =154.17$) constitutes cross-link. Ammonium persulfate (APS) was used as an initiator, and tetramethylethlyenediamine (TEMD) was used as an accelerator of the radical polymerization. Most of produced gels composed of $1.2$ $g$ AA, $6$ $mg$ BIS, $2-15$ $mg$ APS, and $30$ $\mu l$ TEMD, dissolved in $12$ $ml$ deionized water. Oxygen in water was removed by the sufficient degas with an aspirator under ultra-sonic irradiation. Then, pre-gel solution was poured onto a $85$ $mm$ Petri-dish with free upper surface boundary condition. They were left for about $20$ hours under temperature controlled environment in the range of $5-60$ {\degree}. All Petri-dishes were capped in order to suppress water evaporation and keep high humidity. The typical thickness of resultant gel slabs was about $2$ $mm$. Surface patterns were photographed by a digital CCD camera, and those photos were processed by a PC. In fig.\ \ref{fig:pat}, we show typical two types of surface patterns; (a) random, and (b) straight stripe. In these photos, bright places indicate valleys, and dark places indicate bumps, {\it i.e.}, there is vertical roughness on the surfaces. Their characteristic wavelengths are in the order of $mm$. We are going to report what determines the characteristic length scale, and how these patterns are selected. 

\begin{figure}
\begin{center}
\scalebox{0.8}[0.8]{\includegraphics{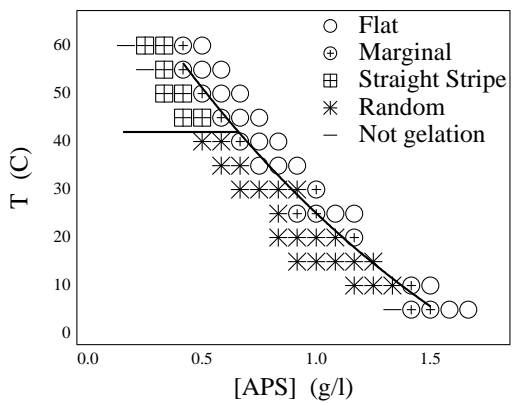}}
\caption{Phase diagram of the pattern formation. Straight stripe patterns are observed in relatively high temperature regime, while random patterns are observed in low temperature regime. }
\label{fig:pd}
\end{center}
\end{figure}

In order to clarify the appearing regime of each surface pattern, we systematically made gel slabs. The obtained phase diagram is shown in fig.\ \ref{fig:pd}. APS concentration $[APS]$ ($g/l$) and temperature $T$ ({\degree}) were chosen as control parameters because surface patterns were sensitive to these two factors. Since mechanical properties of gels and the viscosity of pre-gel solutions usually depend on the concentration of BIS (cross-link), we expected the created patterns also depend on that. However, the concentration of BIS did not affect the structure of this surface patterns very much. In this study, the degree Celsius is used instead of the absolute temperature since the solvent we use is water. Very high $[APS]$ and very high $T$ conditions made uniform flat (no surface pattern) gel slabs. Contrary, insufficient gelation or large scale folded slabs appeared in very low $[APS]$ and very low $T$ conditions. In the middle regime of these limit cases, clear surface patterns were created. That is, surface patterns can be observed in narrow regime of gelation condition. We found that straight stripe patterns tend to appear in relatively high $T$ regime, and random patterns appear in low $T$ regime, as shown in fig.\ \ref{fig:pd}. The mark "Marginal" in fig.\ \ref{fig:pd} means slightly patterned surfaces.

Although the pattern formation with volume phase transition was caused by the mechanical instability, it does not work so effectively in this gelation case. Since the gelation process does not contain drastic volume change, the mechanical instability seems to be less important. Instead, we consider that the diffusion process of the polymer clusters is the most basic dynamics for polymer gelation. Aggregation process (radical chain reaction) is also fundamental for polymer gelation. Therefore, it is natural that the reaction-diffusion coupling causes surface instability. In quasi-two-dimensional space, monomer and polymer clusters diffuse and aggregate inhomogeneously depending on a certain instability. As a result, vertical differences are created owing to the free upper surface boundary condition. In other words, the vertical thickness of gel slabs indicates the degree of concentration of polymer clusters. Therefore, we consider that a two-dimensional reaction-diffusion dynamics is possible to explain this pattern formation.

There are two kinds of reaction-diffusion systems. The one is "activator-inhibitor" system and the other is "substrate-depleted" system. In either case, diffusion coefficients have to satisfy the qualitative requirement $D_v > D_u$ to occur Turing instability. Where $D_v$ is a diffusion coefficient of inhibitor (or substrate), and $D_u$ is that of activator (or depletion). If we assume the substrate-depleted system, the acrylamide monomers and heat respectively correspond to the substrate and the depletion. However, diffusion coefficients conflicts with the requirement in this case. The diffusion of heat is much faster than that of monomers (or polymer clusters). 

On the other hand, the activator-inhibitor system seems to work well. The polymerization speed depends on temperature, and the polymerization itself is exothermic reaction, {\it i.e.}, polymerization has a positive feedback.  This means we can assume the concentration of acrylamide monomers (or polymer clusters) as an activator in this reaction-diffusion system. The polymerization is inhibited by any element or compound that serves as a free radical trap. It is well known that oxygen is such an inhibitor \cite{Chrambach1}. We sufficiently degassed pre-gel solutions to reduce this effect. However, oxygen in air dissolved gradually from the free upper surface boundary of slabs during polymerization. When we made a gel slab under decompressed environment, no surface pattern emerged. This implies the presence of oxygen is essential to create surface patterns. Thus, we regard oxygen as an inhibitor in this reaction-diffusion system. The diffusion coefficient of oxygen in water ($25$ {\degree}) can be approximated as $D_v=2.2 \times 10^{-9}$ $(m^2 /s)$ \cite{Ferrell1}. The diffusion coefficient of acrylamide polymer is expressed as $D_u = 1.24 \times 10^{-8} M_w^{-0.53}$ $(m^2/s)$ \cite{Schwartz1}. Then the requirement $D_v > D_u$ is satisfied even if we substitute the acrylamide monomer molecular weight $M_w=71.08$ to this equation ($D_u=1.29 \times 10^{-9}$ $(m^2 /s)$). Moreover, $D_u$ becomes smaller when polymerization proceeds. 

We subsequently examined the effect of the slab thickness (three dimensional effect). When we made two or three times thicker slabs with the same size Petri-dishes, they did not show clear surface pattern. Due to the exothermic heat of radical polymerization, thermal convection may occur in thick samples. Moreover, the amount of inhibitor oxygen might become insufficient in thick samples because it comes from air. Then, diffusion-induced instability cannot occur. Approximate two-dimensionality is needed to create surface patterns. 

Based on these experimental observations, diffusion of acrylamide monomers (or polymer culsters) and dissolved oxygen are essential to construct the surface patterns with reaction-diffusion dynamics. And the surface patterns are induced by Turing instability.

\begin{figure}
\begin{center}
\scalebox{0.8}[0.8]{\includegraphics{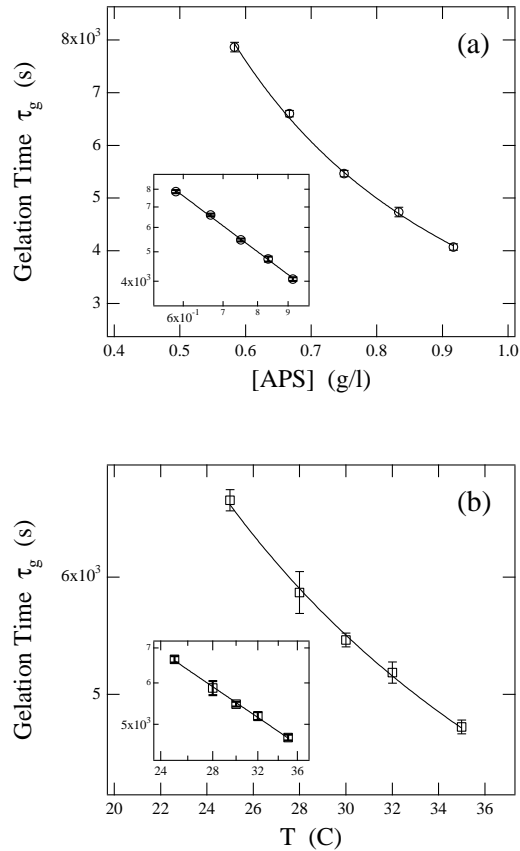}}
\caption{The gelation time as a function of: (a)APS concentration $[APS]$, (b)Temperature $T$. Each plot can be fitted by the forms $\tau_g \sim [APS]^{-\alpha}$ or $\tau_g \sim T^{-\beta}$. We obtained $\alpha=1.46 \pm 0.03$ and $\beta = 1.01 \pm 0.03$. The insets show each log-log plot. }
\label{fig:tau_g}
\end{center}
\end{figure}

The diffusion length determins the wavelength of the instability in reaction-diffusion systems. A characteristic wavelength  of the Turing instability is estimated by the geometric mean of diffusion lengths of activator and inhibitor \cite{Cross1,Walgraef1}. Roughly speaking, the upper limit of the wavelength corresponds to the diffusion length of the inhibitor. The diffusion length  $l_d$ can be estimated as, 
\begin{equation}
l_d \sim (D \tau_g)^{1/2},
\label{eq3}
\end{equation}
where $D$ is a diffusion coefficient and $\tau_g$ is a gelation time. Equation (\ref{eq3}) means that the gelation time should relate to the characteristic wavelength of the surface patterns. Accordingly, we measured the gelation time. We define that the gelation completes when no  pre-gel solution (which does not have yield stress) remains on the Petri-dish. We show the gelation time $\tau_g$ as a function of $[APS]$ in fig.\ \ref{fig:tau_g}(a), and as a function of $T$ in fig.\ \ref{fig:tau_g}(b). The temperature was set to $T=30$ {\degree} in fig.\ \ref{fig:tau_g}(a) experiments, and the concentration of APS was set to $[APS]=0.75$ $g/l$ in fig.\ \ref{fig:tau_g}(b) experiments. In both figures, clearly negative correlation can be confirmed. Solid curves are the fitting curves of 
\begin{equation}
\begin{split}
\tau_g \sim [APS]^{-\alpha}, \\ 
\tau_g \sim T^{-\beta}. 
\end{split}
\label{eq:tau_g}
\end{equation}
We obtained the values $\alpha=1.46 \pm 0.03$ and $\beta=1.01 \pm 0.03$. Although the examined ranges are not so wide owing to the limit of pattern appearance, fittings look very good. 

\begin{figure}
\begin{center}
\scalebox{0.8}[0.8]{\includegraphics{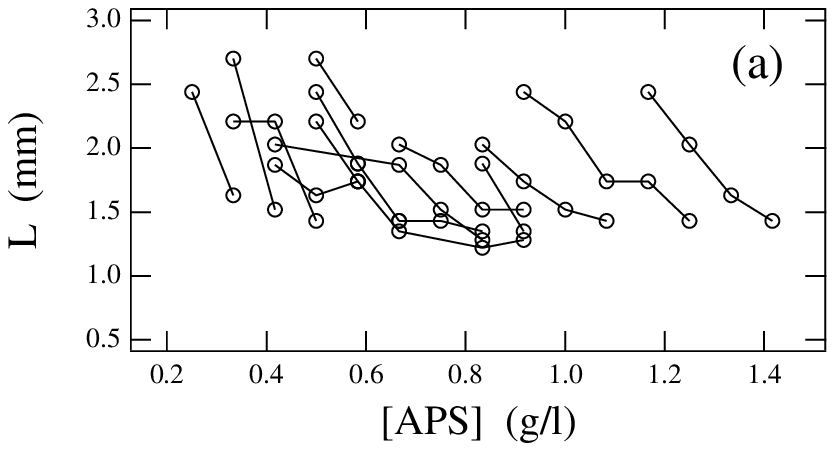}}
\scalebox{0.8}[0.8]{\includegraphics{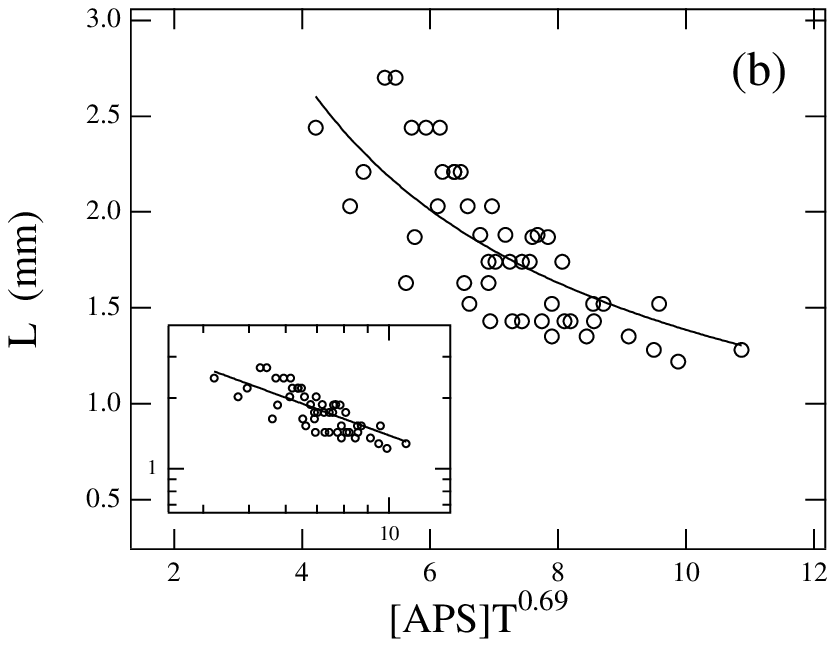}}
\caption{The characteristic length (wavelength) measured by FFT analysis. (a)Data of same temperature are connected. (b)Data are collapsed by the scaled argument of concentration of APS and Temperature. The fitting curve is  $L \sim ([APS]T^{0.69})^{-0.73}$. The inset shows the log-log plot. }
\label{fig:L}
\end{center}
\end{figure}

As shown in fig.\ \ref{fig:pat}, surface patterns seem to have characteristic length scale (wavelength). Hence, we calculated the wavelength using the two-dimensional FFT (Fast Fourier Transform) analysis. Figure \ref{fig:L} shows the measured wavelengths $L$. In fig.\ \ref{fig:L}(a), we show $[APS]$ dependency of the wavelength, and each curve connects the data of same $T$. Negative correlation between $L$ and $[APS]$ can be confirmed.  

Here we discuss the relation between $L$ and $\tau_g$, using scaling exponents $\alpha$ and $\beta$. In general, $D$ is proportional to the absolute temperature  (not the degree Celsius) of solvent \cite{Wilke1}. Thus, the $D$ variation is negligible compared with the $\tau_g$ variation in the measured temperature range, particularly for the scaling discussion. From the eq.\ (\ref{eq3}), we can assume $L$ ($\simeq l_d$) $\sim \tau_g^{1/2}$. It leads to the relations $L \sim [APS]^{-\alpha /2}$ and $L \sim T^{-\beta /2}$ by eqs.\ (\ref{eq:tau_g}). Consequently, we conjecture 
\begin{equation}
L \sim [APS]^{-\alpha /2}T^{-\beta /2} \sim ([APS]T^{\beta /\alpha})^{-\alpha /2},
\label{eq1}
\end{equation} 
because $[APS]$ and $T$ can be controlled independently. On this basis, we tried the data collapse using the eq.\ (\ref{eq1}) scaling. The result with $\alpha /2 =0.73$ and $\beta / \alpha = 0.69$ is shown in fig.\ \ref{fig:L}(b). Whereas the data in fig.\ \ref{fig:L}(b) look a little scattered, this value $\beta / \alpha =0.69$ made a good data collapse rather than other $\beta / \alpha$ values. Namely, these results are fully consistent by means of scaling exponents. In addition, substituting the typical gelation time of our experiments (see fig.\ \ref{fig:tau_g}) $\tau_g = 5000$ ($s$) and $D_v =2.2 \times 10^{-9}$ $(m^2/s)$ into eq.\ (\ref{eq3}), we finally obtain the approximate upper limit of the characteristic wavelength of the instability as $L=3.3$ $(mm)$. This value is certainly close to the upper limit value of the experiment (fig.\ \ref{fig:L}). This fact implies that all quantities have a complete consistency even in coefficients argument, not only in exponents argument. The solid curve in fig.\ \ref{fig:L}(b) is a fitting curve using the values $\alpha /2=0.73$ and $\beta / \alpha=0.69$ ({\it i.e.}, remaining fitting parameter is only the coefficient). 

We have identified the activator and the inhibitor, and discussed the consistency between the wavelength and the diffusion length. They clearly support the diffusion-induced pattern fromation. However, our explanation with Turing instability is partly qualitative. The dynamics we have thought might be too simple. More specific modeling and theoretical study are needed to clarify the dynamics of pattern formation quantitatively. For example, polymer clusters obviously grow step by step during the polymer gelation process. In some condition, polymer clusters grow as fractals and the effective diffusion coefficient changes \cite{Manley1}. Nevertheless, we have discussed the length scale using monomer diffusion. This is only a first order approximation. This might be a reason why the plot in fig.\ \ref{fig:L}(b) scatters. To obtain good statistics in the relation among $L$, $[APS]$, and $T$, more realistic modeling must be taken into account. In particular, reaction terms should be carefully considered to build the model. The gelation reaction might differ from ordinary chemical reactions reported so far in reaction-diffusion systems. That would be a challenging future problem. 

We have mentioned that the straight stripe patterns were observed in high temperature polymerized gel slabs, while random patterns were observed in low temperature regime. Since thermal fluctuation is larger in high temperature regime, one might consider that randomness also should be increased in high temperature regime. However, disordered (random) patterns were observed in low temperature regime, and ordered (straight stripe) patterns were observed in high temperature regime. This sort of pattern selection may be clarified by more detailed numerical simulations or analytic calculations. Moreover, large scale folding is also interesting. Almost all surface patterns we studied in this paper have flat bottom (patterns appeared only on upper surface). However, some slabs have large scale folding structures that include bottom deformations. We could not understand well on this folding problem. This is an open problem. 

When $[APS]$ or $T$ was set larger, surface patterns disappeared abruptly rather than short wavelength patterns. There seems to be the lower limit of $L$. This lower limit conflicts with the scaling of eq.\ (\ref{eq1}). The system might need another condition in order to cause the surface instability. Because of this unknown condition, the appearing regime of surface patterns is very limited. Therefore, the scaled range in fig.\ \ref{fig:L} is narrow (less than one order). This abrupt appearing dynamics should be revealed in more detailed experiment.

Anyway, the instability of the concentration of poly-acrylamide results in uneven surface patterns. Polymer gels are frequently used as substrates of reaction-diffusion systems to control the diffusion coefficient and to suppress the convection effect. Of course, flat and uniform gel slabs have been suitable for many experimental cases such as ordinary reaction-diffusion system and electrophoresis. Therefore, gel slabs with surface patterns have been regarded as wrong samples. They have, however, typical non-equilibrium dissipative structure. We found it and conducted the systematic experiments. It is really easy to create gel slabs with surface patterns. It only needs radical polymerization and its inhibitor like oxygen. This experimental system should be investigated well in future to study reaction-diffusion system.

In summary, we have presented diffusion-induced macroscopic pattern formation on gelation surfaces by the simple experiment. We consider a reaction-diffusion dynamics to describe the pattern formation. We found straight stripe patterns in high temperature regime, while random patterns in low temperature regime. The characteristic wavelength of the surface patterns seems to be approximated by the diffusion length. It relates to the gelation time in terms of scaling exponents. That is, we found the unified scaling function for the wavelength of the patterns. We expect the universality of this scaling among other polymer gelation. In spite of the simplicity of the experiment, it seems to show the diffusion-induced instability, and remains some challenging problems that have great possibility to reveal the reaction-diffusion and gelation dynamics. 

We thank M. Tokita, A. Takada, and H. Honjo for helpful suggestions and discussions. The work was partly supported by a Grant-in-Aid from the Ministry of Education, Science, Sports and Culture of Japan.

\end{document}